\documentclass[aps,pra,twocolumn,superscriptaddress,floatfix,showpacs]{revtex4}

\usepackage{amssymb}
\usepackage{graphicx}

\begin{document}

\title{Simple quantitative estimate of the derivative discontinuity and the
energy gap in atoms and molecules}

\author{F. P. Rosselli}
\affiliation{Divis\~ao de Metrologia de Materiais, Instituto Nacional de 
Metrologia, Normaliza\c{c}\~o e Qualidade Industrial (Inmetro),
Rio de Janeiro 25245-020, Brazil}

\author{A. B. F. da Silva}
\affiliation{Departamento de Qu\'{\i}mica e F\'{\i}sica Molecular\\
Instituto de Qu\'{\i}mica de S\~ao Carlos,
Universidade de S\~ao Paulo,
CP 780, S\~ao Carlos, 13560-970 SP, Brazil}

\author{K. Capelle}
\email{capelle@ifsc.usp.br}
\affiliation{Departamento de F\'{\i}sica e Inform\'atica\\
Instituto de F\'{\i}sica de S\~ao Carlos,
Universidade de S\~ao Paulo,
CP 369, 13560-970 S\~ao Carlos, SP, Brazil}

\date{\today}

\begin{abstract}
The derivative discontinuity of the exchange-correlation functional of 
density-functio\-nal theory is cast as the difference of two types of electron 
affinities. We show that standard Kohn-Sham calculations can be used
to calculate both affinities, and that their difference benefits from
substantial and systematic error cancellations, permitting reliable estimates
of the derivative discontinuity. Numerical calculations for atoms and molecules
show that the discontinuity is quite large (typically several eV), and 
significantly and consistently improves the agreement of the calculated 
fundamental energy gap with the experimental gaps. The percentage error of 
the single-particle gap is reduced in favorable cases by more than one order 
of magnitude, and even in unfavorable cases by about a factor of two.
\end{abstract}


\newcommand{\be}{\begin{equation}}
\newcommand{\ee}{\end{equation}}
\newcommand{\bi}{\bibitem}
\newcommand{\la}{\langle}
\newcommand{\ra}{\rangle}
\renewcommand{\r}{({\bf r})}

\maketitle

From spectroscopy to transport, there is hardly any property of a quantum
many-particle system that does not in some way depend on whether there is a 
gap in the energy spectrum, and what the size of this gap is. The fundamental 
gap is a ground-state property of the $N$-body system, defined
\cite{dftbook,parryang,quantliq} in terms of 
the ground-state energy $E(N)$ as $E_g=[E(N-1)-E(N)]-[E(N)-E(N+1)]$, where 
$E(N-1)-E(N)$ is the energy change upon removing the $N$'th particle from the 
$N$ particle system and $E(N)-E(N+1)$ that upon removing the $N+1$st particle 
from the $N+1$ particle system. In a noninteracting system, this definition 
reduces to the familiar energy gap between single-particle levels. 

For interacting systems, approximate many-body calculations of total energies
and energy gaps are typically performed within the framework of
density-functional theory (DFT) \cite{dftbook,parryang,quantliq} or, 
for small finite 
systems, Hartree-Fock (HF) and post-Hartree-Fock methods. DFT provides, in 
addition to the ground-state density and related quantities, also a set of
single-particle eigenvalues, the so-called Kohn-Sham (KS) spectrum. 
The difference between the energy of the highest occupied and lowest 
unoccupied of these single-particle levels is the KS gap, which
in extended systems become the band structure gap. Neither the KS 
nor the HF single-particle gaps correspond to the experimental gap, 
the former typically being too small and the latter too large. 

Generally, one can write $E_g=E_g^{KS} + \Delta_{xc}$, which defines 
$\Delta_{xc}$ as the difference between the exact fundamental gap and the 
exact KS single-particle gap. In atomic physics and quantum chemistry, the 
importance of a nonzero $\Delta_{xc}$ for chemical hardness is well known 
\cite{parryang,goerling,deproft}. Neglect of $\Delta_{xc}$ has also been shown
to lead to large errors in the calculation of Rydberg excitations
\cite{ryd1,ryd2}, charge-transfer processes and transport
\cite{trans1,trans2,trans3,trans4}, and the ionization probability of atoms 
exposed to strong electromagnetic fields \cite{lein1}. In semiconductors,
approximate energy gaps calculated in DFT often drastically underestimate
the experimental gap \cite{solids1,rubio}. In Mott insulators, in  particular, 
the entire gap is due to $\Delta_{xc}$ \cite{lp,mottepl}.
Recently it was pointed out that a similar discontinuity problem can
also appear in the spin-DFT calculation of spin gaps in materials relevant
for spintronics \cite{nonun}.
The question whether the neglect of $\Delta_{xc}$ or the error in
$E_g^{KS,approx}$ is responsible for the underestimate of the band gap
in solids is considered in a standard textbook in the field to be `{\em of
tremendous practical importance}' \cite{dftbook}, and the calculation of
$\Delta_{xc}$ is ranked in a recent monograph as `{\em certainly one of the
major outstanding problems in contemporary DFT}' \cite{quantliq}, but no
general answer is known.

\begin{table*}
\begin{ruledtabular}
\caption{\label{tab1} Experimental gap, $E_g^{exp}=I^{exp}-A^{exp}$, KS
single-particle gap, $E_g^{KS}$, and KS gap corrected by adding our estimate
for the discontinuity, $E_g^{DFT}=E_g^{KS}+\Delta_{xc}^{est}$, for atoms $Li$
($Z=3$) to $Ca$ ($Z=20$), with exception of the nobel gases.
The values for $\Delta_{xc}^{est}$ were obtained from Eq.~(\ref{estimate})
using the B88LYP functional and the 6-311G(d) basis sets. All data in $eV$.}
\begin{tabular}{c|c|c|c|c|c|c|c|c|c|c|c|c|c|c|c|c|c|c}
Z &3&4&5&6&7&8&9&11&12&13&14&15&16&17&19&20&rms\\
  &Li&Be&B&C&N&O&F&Na&Mg& Al&Si&P&S&Cl&K&Ca&error\\
\hline
$E_g^{KS} $ & 1.40&3.56&0.614&0.598&3.76&0.860&0.820&0.917&3.38& 0.272&0.188&2.01&0.215&0.146&0.601&2.36&7.67\\
$E_g^{DFT}$ & 4.59&9.01&8.12&10.1&14.4&13.2&15.1&4.42&7.19& 5.09&6.43&8.45&7.64&8.88&3.63&5.32&0.606\\
$E_g^{exp}$ & 4.77&9.32&8.02&10.0&14.5&12.2&14.0&4.59&7.65& 5.55&6.76&9.74&8.28&9.36&3.84&6.09&-\\
\end{tabular}
\end{ruledtabular}
\end{table*}

In the present paper we draw attention to an alternative representation of
$\Delta_{xc}$, which casts it as a difference of single-particle eigenvalues, 
similar to the KS gap. We point out
that this relation provides a simple physical interpretation of the elusive
$xc$ discontinuity, and use it to estimate $xc$ discontinuities of atoms and
molecules. The resulting correction $\Delta_{xc}^{est}$ substantially improves 
agreement with experimental fundamental gaps, reducing the percentage deviation
from experiment by more than a factor of $10$ in favorable cases and by 
about a factor of two in  less favorable ones.

The standard representation of $\Delta_{xc}$ is based on ensemble DFT for 
open systems, where all three quantities in $E_g=E_g^{KS} + \Delta_{xc}$ 
can be related to derivative discontinuities of universal density 
functionals \cite{lp,ss,kohn}. The fundamental gap is the derivative 
discontinuity of the total energy
\be
E_g=\left.\frac{\delta E[n]}{\delta n\r}\right|_{N+\eta}
- \left.\frac{\delta E[n]}{\delta n\r}\right|_{N-\eta},
\ee
the KS single-particle gap that of the noninteracting kinetic energy
\be
E_g^{KS}=\left.\frac{\delta T_s[n]}{\delta n\r}\right|_{N+\eta}
- \left.\frac{\delta T_s[n]}{\delta n\r}\right|_{N-\eta},
\ee
and the remaining piece, $\Delta_{xc}$, that of the xc energy
\be
\Delta_{xc}=
\left.\frac{\delta E_{xc}[n]}{\delta n\r}\right|_{N+\eta}
- \left.\frac{\delta E_{xc}[n]}{\delta n\r}\right|_{N-\eta}
=
v_{xc}^+\r - v_{xc}^-\r.
\label{dxcdef}
\ee
In these equations $\eta$ stands for an infinitesimal variation of the 
system's particle number. Equation (\ref{dxcdef}) shows that $\Delta_{xc}$ 
is a system-dependent shift of the $xc$ potential $v_{xc}\r$ as it passes 
from the electron-poor to the electron-rich side of integer $N$. 

Theses three relations are useless to calculate gaps from most currently 
available 
approximate density functionals, which typically have no discontinuities.
For two of the three quantities above, alternative ways of calculation, 
more useful in practice, are widely known. Total energies are calculated 
easily from DFT, so that one can employ the definition of ionization energy 
$I=E(N-1)-E(N)$ and electron affinity $A=E(N)-E(N+1)$ to calculate the 
fundamental gap from
\be
E_g = E(N+1)+E(N-1)-2 E(N) = I - A.
\label{gap}
\ee
Single-particle energies are obtained from the KS equation, as a byproduct of
calculating the total energies, and yield the KS single-particle gap
\be
E_g^{KS}=\epsilon_{N+1}(N)-\epsilon_N(N),
\label{ksgap}
\ee
where $\epsilon_M(N)$ denotes the $M$'th eigenvalue of the $N$-electron system. 
The third term, the $xc$ discontinuity $\Delta_{xc}$, has resisted all attempts
of describing it directly by common density functionals, such as LDA and GGA,
which are continuous as a function of $N$ and thus have no $xc$ discontinuity. 

However, we note that
$I$ and $A$ can be calculated in DFT not only from ground-state energies, 
but also from single-particle eigenvalues, via $I=-\epsilon_N(N)$ and 
$A=-\epsilon_{N+1}(N+1)$ (the analogue of Koopmans' theorem in DFT) 
\cite{dftbook,way}. By using these relations and the definition of 
$E_g^{KS}$ one finds, upon combining Eqs.~(\ref{gap}) and (\ref{ksgap})
\cite{shamschlueter},
\be
E_g= \epsilon_{N+1}(N) - \epsilon_N(N) +\Delta_{xc}
=\epsilon_{N+1}(N+1)-\epsilon_N(N), 
\label{algebra}
\ee
which implies 
\be
\Delta_{xc}=  \epsilon_{N+1}(N+1) - \epsilon_{N+1}(N) = A_{KS}-A. 
\label{estimate}
\ee
In the last step we used the fact that the affinity of the KS system,
$A_{KS}$, is simply the negative of the energy of the lowest unoccupied 
orbital. 

One way to estimate $\Delta_{xc}$ is by subtracting Eq.~(\ref{ksgap}) from 
(\ref{gap}). In the following, we use Eq.~(\ref{estimate}), which is equivalent
in principle, but simpler in practice \cite{footnote}. It also provides an
intuitive interpretation of the discontinuity: in an interacting system, the
electrons repell, hence the energy cost of removing the outermost electron
from the negative species (which is measured by the electron affinity) is
reduced, $A<A_{KS}$, and a positive $\Delta_{xc}$ results.

If the right-hand side of Eq.~(\ref{estimate}) could be calculated exactly,
this procedure would determine the exact $xc$ discontinuity. An estimate of
$\Delta_{xc}$ is thus obtained by using in (\ref{estimate}) the KS eigenvalues 
obtained in two approximate KS calculations, one for the neutral species, the 
other for the anion. 

Such approximate calculations involve two distinct types of errors, one
associated with the approximations used for the $xc$ functional, the other 
with the finite size of the basis set. As a consequence, each of the two
affinities in Eq.~(\ref{estimate}) is predicted wrongly by standard 
combinations of functionals and basis sets. Typically, the self-interaction 
error inherent in common LDA and GGA type functionals shifts the eigenvalues 
up, in some cases so much that the anions become unbound \cite{trickey}. 
On the other hand, the finiteness of the basis set artificially stabilizes 
the anion \cite{trickey}. As a consequence of this error cancellation, 
practical methods for calculating affinities from LDA and GGA are available 
\cite{trickey,gals,geerlings,vydrov}. 

Our key argument is that derivative discontinuities are protected from 
functional errors and basis-set errors by a distinct additional error 
compensation, independent of the one just described.
Namely, Eq.~(\ref{estimate}) casts $\Delta_{xc}$ as a difference of two
affinities. Even if each is predicted wrongly on its own, their difference
is expected to benefit from substantial additional error cancellation.
In fact, if all KS eigenvalues are shifted up by roughly the same amount,
energy differences are preserved, and even positive eigenvalues (unbound
anions) can provide reasonable discontinuities for the bound neutral system.

We call this calculation of $\Delta_{xc}$ by means of 
Eq.~(\ref{estimate}) an estimate, and not an approximation, to stress that 
it exploits an error cancellation that is hard to quantify {\em a priori}. 
However, in recent work on models of harmonically confined systems 
\cite{confined} this estimate was found to lead to significantly improved
gaps. Here we explore the performance of Eq.~(\ref{estimate}) in {\em ab 
initio} calculations for atoms and molecules. 

In Table~\ref{tab1} we compare, for 16 light atoms, the experimental gap, the 
KS single-particle gap, and the DFT gap, defined as the sum of the KS gap and 
the estimated $xc$ discontinuity. The KS calculations were performed with the 
{\tt GAUSSIAN 98} \cite{gaussian} program, using the B88-LYP functional and 
the $6-311G(d,p)$ basis sets. Table~\ref{tab1} shows that the error of 
the KS gap is significantly and consistently reduced by adding the estimated 
$xc$ discontinuity to the KS gap, dropping by more than an order of magnitude 
--- from $7.67 eV$ to $0.606 eV$ --- over the data set in Table~\ref{tab1}. 
This large drop, together with the fact that the improvement 
is systematic (obtained not only on average, but in every individual case), 
strongly suggests that Eq.~(\ref{estimate}) is a reliable and useful way of 
obtaining the discontinuity. 

\begin{figure}[t!]
\includegraphics[height=60mm,width=70mm,angle=0]{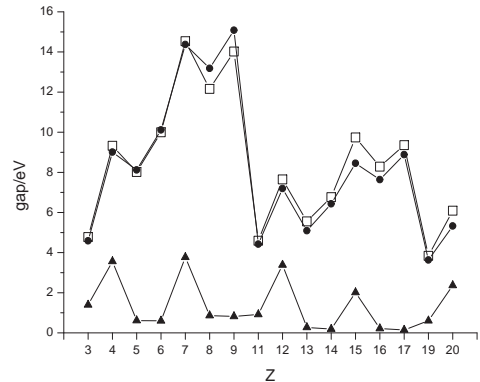}
\caption{\label{fig1} Plot of the data in Table \ref{tab1}:
Kohn-Sham single-particle gap (triangles), experimental fundamental gap (open
squares) and Kohn-Sham single-particle gap corrected by adding our estimate 
of the discontinuity (full circles), for 16 light atoms. The lines are 
guides for the eye, illustrating that $\Delta_{xc}^{est}$ recovers not only
the overall value, but also fine details of the behaviour as a function of $Z$ that were lost in the KS gap.}
\end{figure}

Figure \ref{fig1} is a plot of the data in Table~\ref{tab1}, revealing
that $\Delta_{xc}$ roughly follows the
atomic shell structure. Particularly small discontinuities are found for 
atoms with one electron outside a closed shell, such as $Li$ and $Na$. 
The largest discontinuities are, however, not observed for closed-shell 
systems but for systems one electron short of a closed shell, as is seen 
comparing $F$ with $Ne$ or $Cl$ with $Ar$. We interpret this by means of
Eq.~(\ref{estimate}) as a consequence of the fact that $\Delta_{xc}$ is 
related to two affinities, which involve negative species with 
one additional electron, leading to a closed shell for $F^-$ and $Cl^-$.

Specifically for the $Be$ atom, we can further compare with independent
theoretical expectations, because the discontinuity of this atom has previously
been estimated by Jones and Gunnarsson (JG) \cite{gunjones} by comparing the
experimental gap to a near-exact KS gap obtained earlier by Pedroza and 
Almbladh \cite{pedroza} from CI densities and approximate inversion of the KS 
equation. Our value $\Delta_{xc}^{Be}=5.5eV$ is encouragingly close to
$\Delta_{xc}^{Be,JG}=5.7 eV$.

Next, we turn to molecules. In Table~\ref{tab2} we compare our estimate of 
$\Delta_{xc}$ to many-body values of $\Delta_{xc}$ and to 
experimental fundamental gaps. The many-body discontinuity is obtained 
\cite{tozer} by performing coupled-cluster calculations to generate a 
near-exact density, followed by inversion of the KS equation to obtain the 
corresponding near-exact 
KS potential, solution of the KS equation with that potential to obtain the 
near-exact KS gap, and subtraction of that gap from the experimental 
fundamental gap \cite{tozer}. The first step is impractical for larger 
systems, whereas the last step involves using the experimental gap, which 
makes the method empirical. For these reasons, the simple estimate 
obtained from Eq.~(\ref{estimate}) may constitute a useful alternative, 
provided it turns out to be sufficiently reliable.

In fact, the estimated value of $\Delta_{xc}^{est}$ depends on the chosen basis
set and functional, as well as on whether the anion geometry is separately 
optimized (leading to adiabatic affinities and discontinuities) or hold fixed
at that of the neutral species (vertical affinities and discontinuities).
Tests of different combinations of methodologies indicate that best 
({\em i.e.}, closest to experiment) gaps are obtained if the discontinuity 
is calculated from vertical affinities. 

\begin{table*}
\begin{ruledtabular}
\caption{\label{tab2} Comparison of calculated and experimental gaps for 
small molecules. First column: experimental gap. Second column: KS
single-particle gap. Third column: percentage deviation of KS gap from
experimental gap. Fourth column: derivative discontinuity estimated from
single-particle eigenvalues, obtained from the B88-LYP functional on the
6-311G(d,p) basis sets. Fifth column: DFT gap. Sixth column: percentage 
deviation of DFT gap from experimental gap. Seventh column: many-body 
estimate of the derivative discontinuity. All values in eV.}
\begin{tabular}{c|c|c|c|c|c|c|c}
system & 
$E_g^{expt.}$, Ref.~\cite{pearson} &
$E_g^{KS}$ &
$\%$ dev. &
$\Delta_{xc}^{est}$, Eq.(\ref{estimate}) &
$E_g^{DFT} = E_g^{KS} + \Delta_{xc}^{est}$ &
$\%$ dev. &
$\Delta_{xc}$, Ref.~\cite{tozer} \\
\hline
$CO$ &15.8&7.05&-55.4&9.04&16.09&1.83&8.44\\
$H_2CO$ &12.4&3.66&-70.5&8.31&11.97&-3.43&8.16\\
$H_2S$ &12.6&5.68&-54.9&5.33&11.01&-12.6&6.53\\
$HCN$ &15.9&8.06&-49.3&8.24&16.30&2.53&7.89\\
$N_2$ &17.8&8.24&-53.7&9.74&17.98&1.15&9.25\\
$PH_3$ &11.9&6.51&-45.3&5.25&11.76&-1.18&5.99\\
$Cl_2$ &9.20&2.38&-74.1&7.24&9.620&4.57&-\\
$SO_2$ &11.2&3.25&-71.0&7.94&11.19&-0.09&-\\
$C_2H_4$ &12.3&5.76&-53.2&7.10&12.86&4.57&6.53\\
$C_2H_2$ &14.0&6.97&-50.2&5.34&12.32&-12.0& 7.08\\
$H_2O$ & 19.0&6.47&-66.0&6.15&12.61&-33.6&11.4\\
$NH_3$ & 16.3&6.01&-63.1&5.56&11.58&-29.0&10.1\\
$HF$ &22.0&8.63&-60.8&6.82&15.45&-29.8&11.7\\
$CH_4$ &20.5&10.3&-49.9&4.88&15.15&-26.1&11.4\\
\end{tabular}
\end{ruledtabular}
\end{table*}

Table~\ref{tab2} shows that the systems
fall in two classes. For one class, comprising $CO$, $H_2CO$, $H_2S$, $HCN$, 
$N_2$, $PH_3$, $Cl_2$, $SO_2$, and $C_2H_4$, the gap error with 
respect to experiment is reduced by a similar margin as for atoms, or even 
more, and estimated and calculated $\Delta_{xc}$ agree well.
For the other class, comprising $H_2O$, $NH_3$, $HF$, $CH_4$ and perhaps
$C_2H_2$, the percentage
error of the gap drops by a factor of two, instead of by one order of magnitude,
and the estimate recovers about 50\% of the many-body value of $\Delta_{xc}$.
An empirical indicator of which class a system belongs to is the sign of the
KS affinity: if this is negative ({\em i.e.}, the lowest unoccupied KS orbital 
has positive eigenvalue) the system belongs to class II. We note that this is 
not a stability criterium because it employs the (unphysical) KS affinity and
because for all molecules in Table~\ref{tab2} (with exception of $Cl_2$ and 
$SO_2$) the anionic species is experimentally unstable. Rather, it indicates 
a partial loss of the error cancellation on which our use of
Eq.(\ref{estimate}) is based. Nevertheless, even in these "unfavorable" 
cases, the estimate still provides a systematic improvement on the KS gap,
reducing the error with respect to experiment by a factor of two. For both
classes, $\Delta_{xc}^{est}$ clearly provides a quantitative correction to
single-particle gaps, which may be useful in improving, e.g., the DFT 
description of chemical hardness \cite{goerling,deproft} or of transport 
through single molecules \cite{trans1,trans2}.



In summary, we have cast the derivative discontinuity of DFT as a difference
of two affinities, Eq.~(\ref{estimate}), and used approximate KS calculations 
of these to obtain estimates for the discontinuity in atoms and molecules. 
Our results are consistent with previous results, where available,
and significantly and consistently reduce the error between calculated and 
measured fundamental gaps. 

This work was sup\-por\-ted by FAPESP, CAPES and CNPq.

\end{document}